\documentclass[a4paper,11pt]{article}
\pdfoutput=1
\usepackage{jcappub}
\usepackage{mathtools}

\graphicspath{{plots/}}

\newcommand{\lhs}{\lambda_{hs}}
\newcommand{\mdm}{m_s}
\newcommand{\rR}{\rho_R}
\newcommand{\rp}{\rho_\phi}
\newcommand{\Gp}{\Gamma_\phi}
\newcommand{\gs}{g_\star}

\newcommand{\sv}{\langle\sigma v\rangle}
\newcommand{\Tini}{T_\text{i}}
\newcommand{\Tc}{T_\text{c}}
\newcommand{\Tend}{T_\text{end}}
\newcommand{\aini}{a_{\rm i}}
\newcommand{\ac}{a_\text{c}}
\newcommand{\aend}{a_\text{end}}

\title{From WIMPs to FIMPs:\\Impact of Early Matter Domination}

\author[a]{Javier Silva-Malpartida,} \emailAdd{javier.silvam@pucp.edu.pe}
\author[b]{Nicolás Bernal,} \emailAdd{nicolas.bernal@nyu.edu}
\author[a]{\\Joel Jones-Pérez,} \emailAdd{jones.j@pucp.edu.pe}
\author[c]{and Roberto A. Lineros} \emailAdd{roberto.lineros@ucn.cl}

\affiliation[a]{Sección Física, Departamento de Ciencias, Pontificia Universidad Católica del Perú\\
Apartado 1761, Lima, Peru.}
\affiliation[b]{New York University Abu Dhabi\\
PO Box 129188, Saadiyat Island, Abu Dhabi, United Arab Emirates.}
\affiliation[c]{Departamento de Física, Universidad Católica del Norte\\
Avenida Angamos 0610, Casilla 1280, Antofagasta, Chile.}

\abstract{
In the context of non-standard cosmologies, an early matter-dominated (EMD) era can significantly alter the conventional dark matter (DM) genesis. In this work, we reexamine the impact of an EMD on the weakly- and feebly-interacting massive particle (WIMP and FIMP) paradigms. EMD eras significantly modify the genesis of DM because of the change in the Hubble expansion rate and the injection of entropy. The WIMP paradigm can be realized with couplings much smaller than in the standard cosmological scenario, whereas much larger couplings are required in the FIMP case. Using the singlet-scalar DM model as a case study, we show that these results can lead to a continuous transition between the WIMP and FIMP scenarios, with results that are also applicable to other DM models. This broadens the parameter space consistent with observed DM levels and suggests that even elusive FIMP scenarios may be within the reach of future experimental searches.}

\begin{document}
\maketitle

\section{Introduction}
\label{sec:intro}
There is compelling evidence for the existence of dark matter (DM), a non-baryonic component of the Universe. This unknown form of matter has a significant abundance, exceeding the amount of ordinary baryonic matter by a factor of five~\cite{Planck:2018vyg, Cirelli:2024ssz}. The most commonly assumed production mechanism for DM in the early Universe is the freeze-out paradigm, with a WIMP (weakly interacting massive particle) as a candidate for DM. In this framework, DM interacts with the standard model (SM) particles of the thermal plasma, achieving thermal equilibrium in the early Universe. As the Universe expands and cools, these interactions become less frequent, leading to thermal decoupling also known as freeze-out process, which accounts for the observed DM relic abundance today~\cite{Arcadi:2017kky, Roszkowski:2017nbc, Arcadi:2024ukq}. To align with observations, a thermally averaged annihilation cross-section $\mathcal{O}(10^{-26})$~cm$^3$/s is generally needed~\cite{Steigman:2012nb}. In the past decade, this mechanism gained significant popularity within the scientific community. However, negative results from various experiments, including direct and indirect detection, as well as collider searches, have greatly weakened support for this paradigm and forced the community to look for other horizons.

From a particle physics perspective, other DM production mechanisms exist. For example, DM can be generated from the SM bath by nonthermal processes, as in the case of freeze-in, with FIMPs (feebly interacting massive particles) as candidates for DM~\cite{McDonald:2001vt, Choi:2005vq, Kusenko:2006rh, Petraki:2007gq, Hall:2009bx, Elahi:2014fsa, Bernal:2017kxu}. Very suppressed interaction rates between the dark and visible sectors, required to avoid thermalization, can occur in the context of renormalizable portals with couplings of the order $\mathcal{O}(10^{-11})$, as in the case of infrared (IR) FIMPs. Alternatively, ultraviolet (UV) FIMPs appear in the context of nonrenormalizable interactions, suppressed by a large energy scale~\cite{Elahi:2014fsa}, typically higher than the maximal temperature reached by the SM thermal bath.

Interestingly, from a cosmological perspective, there is also a vast margin to play with. The so-called standard cosmology is sustained by the usual assumptions: $i)$ the SM entropy was conserved, and $ii)$ the Hubble expansion rate of the Universe was dominated by SM radiation; both valid from the end of inflationary reheating until the onset of Big Bang nucleosynthesis (BBN) at $t \sim 1$~s. Nevertheless, even though this picture is appealing due to its minimalistic approach, there is actually no strong observational evidence to support it, and there exist many reasonable scenarios leading to departures from these assumptions (see Ref.~\cite{Allahverdi:2020bys} for a review). For example, the existence of a heavy long-lived particle could lead to an early matter domination (EMD) era, violating both of the assumptions above. Taking into account that the energy density of nonrelativistic matter scales as $a^{-3}$ (with $a$ being the cosmic scale factor of the Universe) while the energy density of free radiation scales as $a^{-4}$, one can argue that the former could eventually dominate over the latter even if initially subdominant, provided that the massive particle is sufficiently long-lived. In this case, the only strong constraint on the long-lived heavy particle is that it must decay before the onset of BBN, that is, the EMD period must end before BBN starts, implying an early injection of entropy into the SM plasma.

DM production during EMD eras has been intensively studied in the literature, usually triggered by a long-lived massive particle~\cite{Giudice:2000ex, Fornengo:2002db, Pallis:2004yy, Gelmini:2006pw, Drees:2006vh, Yaguna:2011ei, Roszkowski:2014lga, Drees:2017iod, Bernal:2018ins, Bernal:2018kcw, Cosme:2020mck, Arias:2021rer, Bernal:2022wck, Bhattiprolu:2022sdd, Haque:2023yra, Ghosh:2023tyz, Silva-Malpartida:2023yks, Barman:2024mqo}, or by primordial black holes~\cite{Green:1999yh, Khlopov:2004tn, Dai:2009hx, Fujita:2014hha, Allahverdi:2017sks, Lennon:2017tqq, Morrison:2018xla, Hooper:2019gtx, Chaudhuri:2020wjo, Masina:2020xhk, Baldes:2020nuv, Gondolo:2020uqv, Bernal:2020kse, Bernal:2020ili, Bernal:2020bjf, Bernal:2021akf, Cheek:2021odj, Cheek:2021cfe, Bernal:2021yyb, Bernal:2021bbv, Bernal:2022oha, Cheek:2022dbx, Mazde:2022sdx, Cheek:2022mmy}. However, it can also modify baryogenesis~\cite{Davidson:2000dw, Giudice:2000ex, Allahverdi:2010im, Beniwal:2017eik, Allahverdi:2017edd, Bernal:2017zvx, Chen:2019etb, Bernal:2022pue, Chakraborty:2022gob} or the expected spectrum of primordial gravitational waves~\cite{Assadullahi:2009nf, Durrer:2011bi, Alabidi:2013lya, DEramo:2019tit, Bernal:2019lpc, Figueroa:2019paj, Bernal:2020ywq}. In this paper, we investigate the genesis of DM in the early Universe, during a period of EMD. In particular, we focus on the WIMP and IR FIMP production mechanisms and their variations owing to the changes in the Hubble expansion rate and the injection of entropy. As a working example, we consider the scalar singlet DM model~\cite{Silveira:1985rk, McDonald:1993ex, Burgess:2000yq}, where DM is stable due to $\mathbb{Z}_2$ parity and communicates with the SM through the Higgs portal. Interestingly, we find that in the context of EMD scenarios, there is a continuous and smooth transition between the WIMP and FIMP solutions that greatly broadens the parameter space that fits the observed DM abundance. Large regions of the favored parameter space could be probed by next-generation experiments, even in the case of the elusive FIMP scenario.

The manuscript is organized as follows. In Section~\ref{sec:setup}, we introduce EMD scenarios and the production of DM through the WIMP and FIMP mechanisms in this non-standard cosmological setup. Then, in Section~\ref{sec:DM_density}, we characterize and study the role of the coupling portal strength and the effect of EDM in determining the relic density and DM nature, in the WIMP and FIMP paradigms. Finally, the conclusions are given in Section~\ref{sec:concl}.

\section{Early matter-dominated Universe scenario} \label{sec:setup}
In the period between the end of inflationary reheating and the matter-radiation equality, we assume a Universe with an energy density dominated by SM radiation and a non-relativistic field $\phi$. The corresponding Boltzmann equations for the evolution of the energy densities $\rp$ and $\rR$ for $\phi$ and the SM, respectively, are given by
\begin{align}
    \frac{d\rp}{dt} + 3\, H\, \rp &= - \Gp\, \rp\,, \label{boltz_phi} \\
    \frac{d\rR}{dt} + 4\, H\, \rR &= + \Gp\, \rp\,. \label{boltz_R}
\end{align}
Here, $\Gp$ is the total decay width of $\phi$ into radiation,\footnote{We disregard possible direct $\phi$ decays into DM particles $s$ with mass $\mdm$, which is a good assumption as long as the corresponding branching fraction Br$_{\phi \to s s} \lesssim 10^{-4}~\mdm/(100~\text{GeV})$~\cite{Drees:2017iod, Arias:2019uol}.} and $H$ is the Hubble expansion rate, given by the Friedmann equation
\begin{equation} \label{eq:hub}
    H^2 = \frac{\rR + \rp}{3\, M_P^2}\,,
\end{equation}
with $M_P \simeq 2.4 \times 10^{18}$~GeV being the reduced Planck mass. The SM energy density is defined as a function of the temperature $T$ of the SM photons as
\begin{equation} \label{eq:rad_temp}
    \rR(T) = \frac{\pi^2}{30}\, \gs(T)\, T^4,
\end{equation}
with $\gs(T)$ being the effective number of relativistic degrees of freedom contributing to the energy density $\rR$~\cite{Drees:2015exa}.

For DM production out of the SM bath, the evolution of the DM number density $n_s$ can be studied with the transport Boltzmann equation
\begin{equation} \label{eq:boltzdm}
    \frac{dn_s}{dt} + 3\, H\, n_s = - \sv \left(n_s^2 - n_\text{eq}^2\right),
\end{equation}
where $\sv(T)$ is the 2-to-2 thermally-averaged cross-section for the pair annihilation of DM particles into a couple of SM states, and $n_\text{eq}(T)$ corresponds to the equilibrium DM number density. For non-relativistic DM particles of mass $\mdm$ with a single internal degree of freedom
\begin{equation}
    n_{\rm eq}(T) \simeq \left(\frac{\mdm\, T}{2\pi}\right)^{3/2} e^{-\mdm/T}.
\end{equation} 

In Eq.~\eqref{eq:boltzdm}, two rates compete: the production rate $\Gamma \equiv \sv\, n_{\rm eq}$ between the dark and visible sectors, and the expansion rate $H$. If at some point we have $\Gamma \gg H$, the two sectors equilibrate and DM is called a WIMP. The departure from chemical equilibrium occurs at a temperature $T_\text{fo}$ that can be estimated by the equality
\begin{equation} \label{eq:fo_cond}
    \left. \frac{\Gamma}{H}  \right|_{T = T_{\rm fo}} \simeq  1\,.
\end{equation}
In the usual freeze-out scenario $T_{\rm fo} \simeq \mdm/25$, with a small logarithmic mass dependence. Alternatively, if for all temperatures $\Gamma \ll H$, DM never reaches chemical equilibrium with the SM model. This nonthermal mechanism is known as freeze-in. Even if DM is gradually generated throughout the history of the early Universe, the peak of its production occurs at a temperature $T_{\rm fi}$ which typically corresponds to the maximum between the DM mass and the mediator mass.

In order to agree with the total observed DM relic density, the asymptotic value of the DM yield at low temperatures $Y_0 \equiv n_0/s_0$, where $n_0$ is the present DM number density and $s_0 \simeq 2.69 \times 10^3$~cm$^{-3}$ is the present entropy density~\cite{ParticleDataGroup:2022pth}, must satisfy
\begin{equation} \label{eq:presentY0}
    \mdm\, Y_0 = \frac{\Omega h^2\, \rho_c}{s_0\, h^2} \simeq 4.3 \times 10^{-10}~\text{GeV},
\end{equation}
where $\rho_c \simeq 1.05 \times 10^{-5}~h^2$~GeV/cm$^3$ is the critical energy density of the Universe and $\Omega h^2 \simeq 0.12$ is the observed DM relic abundance~\cite{Planck:2018vyg}. 

\subsection{Stages in the evolution of the early Universe} \label{sec:EMD}
As commented previously, here we assume that at some point in the history of the Universe an EMD era existed. The evolution of the background can be followed by solving Eqs.~\eqref{boltz_phi} and~\eqref{boltz_R}, this can be done by analytical approximations followed by a full-numerical solution for consistency. However, before this, we divide the evolution into four characteristic stages, which will be explained below.

In the first stage ({\bf Stage~1}), $\phi$ is subdominant with respect to the SM, and therefore the Universe follows a standard expansion era driven by $\rR$ with conservation of the SM entropy, which implies that $\rR(a) \propto a^{-4}$, $H(a) \propto a^{-2}$ and $T(a) \propto a^{-1}$. However, the energy density of $\phi$ continuously increases as $\rp(a) \propto a^{-3}$ with respect to $\rR$, as it is a non-relativistic species. It is important to note that the BICEP/Keck bound on the tensor-to-scalar ratio implies an upper bound on $H < H^{\rm CMB}_I = 2.0 \times 10^{-5}~M_P$~\cite{BICEP:2021xfz}. This can be used to derive an upper limit on the maximum temperature reached by the SM plasma $T \lesssim 2.5 \times 10^{15}$~GeV under the assumption of instantaneous reheating~\cite{Barman:2021ugy}, and therefore the production from the SM plasma of heavier particles.

The second stage ({\bf Stage~2}) starts at a temperature $T = \Tini$, defined by the equality $\rR(\aini) \equiv \rp(\aini)$, where $\aini \equiv a(\Tini)$ is the corresponding scale factor. As in this period, the Hubble expansion rate is dominated by $\phi$, $H(a) \propto \sqrt{\rp(a)} \propto a^{-3/2}$. In this era $\phi$ is not effectively decaying, so the SM entropy is still conserved and hence the SM radiation is free: $\rR(a) \propto a^{-4}$ and $T(a) \propto a^{-1}$.

{\bf Stage~3} starts at $T = \Tc$, when $\phi$ begins to effectively decay into SM particles. As SM is sourced, instead of scaling as free radiation, it scales as $\rR(a) \propto a^{-3/2}$, which implies that the SM temperature scales as $T(a) \propto a^{-8/3}$. In this nonadiabatic era, the entropy of the SM is not conserved. Additionally, the Universe is still dominated by $\phi$ and then $H(a) \propto a^{-3/2}$ is still valid. Stage~3 ends at a temperature $\Tend$ that can be estimated by the equality $\Gp \equiv H(\Tend)$, corresponding to
\begin{equation} \label{eq:trh_cond}
    \Gp= \frac{\pi}{3}\, \sqrt{\frac{\gs(\Tend)}{10}}\, \frac{\Tend^2}{M_P}\,.
\end{equation}
To avoid spoiling the success of BBN, the end of Stage~3 must satisfy $\Tend > T_\text{BBN} \simeq 4$~MeV~\cite{Sarkar:1995dd, Kawasaki:2000en, Hannestad:2004px, DeBernardis:2008zz, deSalas:2015glj}. Stages~2 and~3 are usually referred to as the EMD era, the former adiabatic, while the latter nonadiabatic.

In the final stage ({\bf Stage~4}), the standard cosmological evolution of the Universe is recovered because $\phi$ decays exponentially fast ($\rp \propto e^{-\Gp/H}$). The Hubble expansion rate is driven by SM radiation, and the SM entropy is again conserved.

All in all, the evolution of the Hubble expansion rate can be approximated as
\begin{equation} \label{eq:scalingH}
    H(a) \propto
    \begin{dcases}
        a^{-2}   & \text{for } a < \aini \hspace{15mm}\text{(Stage~1)},\\
        a^{-3/2} & \text{for } \aini < a < \ac \hspace{6.5mm}\text{(Stage~2)},\\
        a^{-3/2} & \text{for } \ac < a < \aend \hspace{3mm}\text{(Stage~3)},\\
        a^{-2} & \text{for } \aend < a \hspace{11.6mm}\text{(Stage~4)},
    \end{dcases}
\end{equation}
while the temperature of the SM bath is
\begin{equation} \label{eq:scalingT}
    T(a) \propto
    \begin{dcases}
        a^{-1}   & \text{for } a < \aini \hspace{15mm}\text{(Stage~1)},\\
        a^{-1} & \text{for } \aini < a < \ac \hspace{6.5mm}\text{(Stage~2)},\\
        a^{-3/8} & \text{for } \ac < a < \aend \hspace{3mm}\text{(Stage~3)},\\
        a^{-1} & \text{for } \aend < a \hspace{11.6mm}\text{(Stage~4)}.
    \end{dcases}
\end{equation}

\begin{figure}[t]
    \centering
    \includegraphics[width=0.44\textwidth]{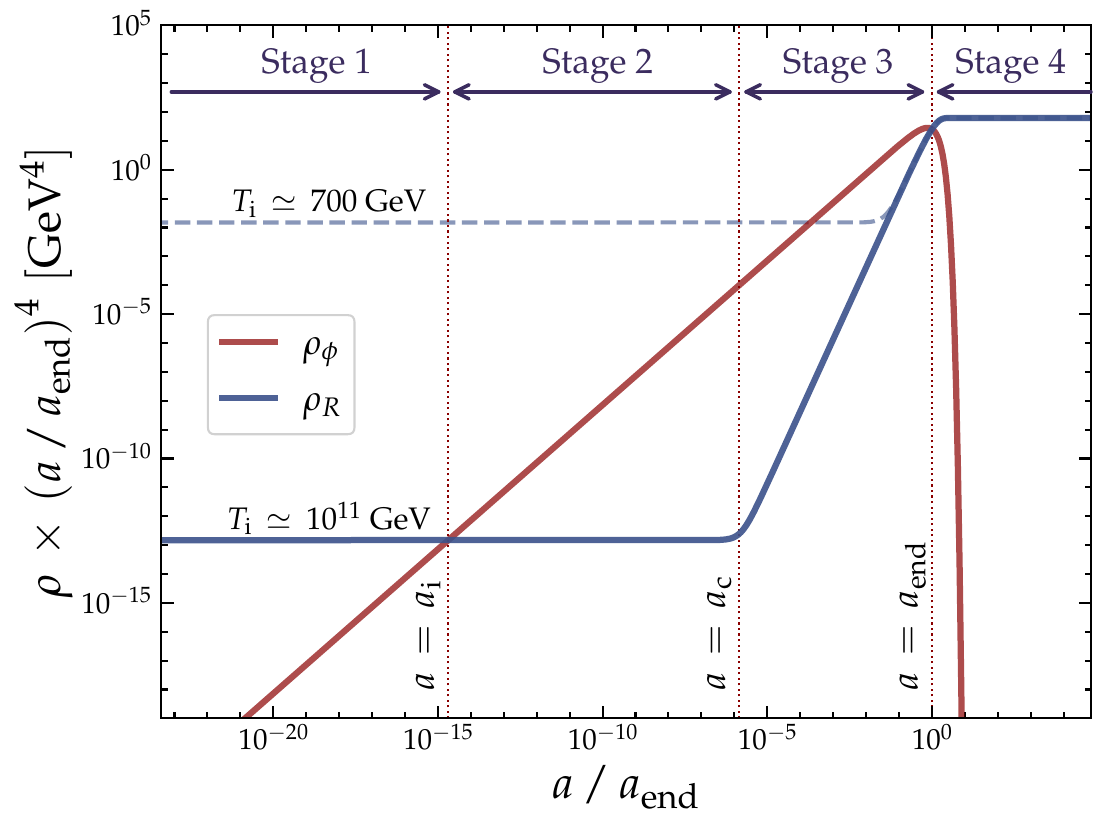}\qquad
    \includegraphics[width=0.495\textwidth]{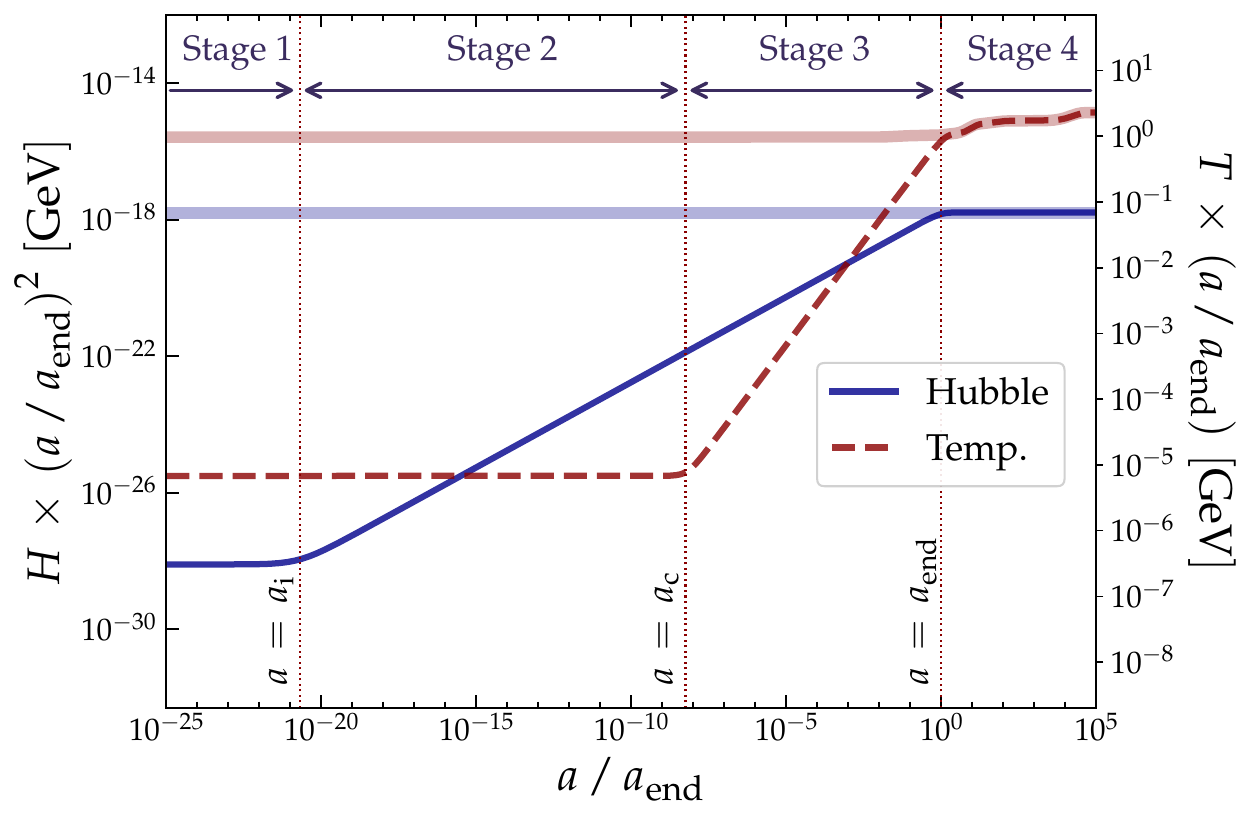}
    \caption{Left: Evolution of the energy densities of SM radiation $\rR$ and matter $\rp$ for two different values of $\Tini$ (left panel) and the Hubble parameter $H$ and temperature $T$ of the SM radiation bath (right panel) as a function of the scale factor $a$. Right: The faint lines represent the standard scenario dominated by SM radiation, for $\Tini \simeq 10^{11}$~GeV. In both panels $\Tend = 1$ GeV. The values for the scale factor $a = \aini$, $\ac$ and $\aend$ are overlaid and meant for $\Tini \simeq 10^{11}$~GeV.}
    \label{fig:den1}
\end{figure}

Having understood the evolution of the background analytically, we now solve it numerically, using the $4^\text{th}$-order \texttt{Runge-Kutta} method, to obtain the evolution of $\rp$ and $\rR$ as a function of the scale factor $a$. The results are shown in Fig.~\ref{fig:den1}. The left panel shows the evolution of $\rp$ (solid red line) and $\rR$ (solid blue line), as a function of the scale factor $a$, for a specific value of $\Tend =1$~GeV, and two values of $\Tini$. Note that the curves have been normalized with the scaling corresponding to free radiation. As expected, $\rp(a)\times a^4 \propto a$ during Stages~1 to~3, and is exponentially suppressed during Stage~4. In turn, $\rR(a)\times a^4$ deviates from flat in Stage~3, where the entropy injection is sizable. An important point to note is that the entropy injection that characterizes Stage~3 brings $\rR$ to the expected value at $\Tend$. Thus, it can be understood that the entropy injected for the curve with $\Tini=10^{11}$~GeV will be higher than that for the curve with $\Tini=700$~GeV. Thus, having fixed $\Tend$, one can view $\Tini$ as a measure of the change in entropy in the system.

The right panel of Fig.~\ref{fig:den1} shows the evolution of $T$ (dashed red line) and $H$ (solid blue line), taking $\Tend=1$~GeV and fixing $\Tini \simeq 10^{15}$~GeV. Both are independently normalized, so they appear flat\footnote{Flat up to changes of the relativistic number of degrees of freedom. An observable bump is seen at $T \sim 0.3$~GeV, due to the QCD phase transition, which leads to a sudden decrease in the effective relativistic degrees of freedom in the thermal plasma.} when following the scaling of standard cosmology (shown in faint red and blue, respectively). We also show the four aforementioned stages, with boundaries in $a=\aini$, $\ac$, $\aend$. The scaling of $H$ and $T$ follows the expected behavior described in Eqs.~\eqref{eq:scalingH} and~\eqref{eq:scalingT}, respectively.

\subsection{Dark matter genesis}
The evolution of the DM number density is obtained by solving Eq.~\eqref{eq:boltzdm}. To this end, we use the code originally developed in Ref.~\cite{Silva-Malpartida:2023yks}, which solves the Boltzmann equation by using an \texttt{implicit Euler} technique, spanning from the early stages to the current state of the Universe.\footnote{We note that other codes that solve the evolution of the DM density in the context of non-standard cosmologies exist (although they do not explore the reheating scenario), such as the ones presented in Refs.~\cite{Dutra:2021phm, Cheek:2022dbx, Cheek:2022mmy, Karamitros:2023uak}.} The thermally-averaged cross-section $\sv$ is evaluated using the standard relation~\cite{Gondolo:1990dk}
\begin{equation} \label{sigv}
    \sv(T) = \int_{4\mdm^2}^{\infty} \mathrm{d}s\, \frac{(s-4\mdm^2) \, \sqrt{s} \, K_{1}\left ( \sqrt{s}/T \right ) \sigma(s)}{8 \,T\, \mdm^{4}\, K_2^2\left ( \mdm/T \right ) }\,,
\end{equation}
where $K_i$ is the modified Bessel function of the second kind of $i^\text{th}$ order. Given a total DM annihilation cross-section $\sigma(s)$, we perform a numerical integration of Eq.~\eqref{sigv} using the \texttt{adaptive Simpson's} method and obtain $\sv$ as a function of the temperature $T$ of the SM radiation plasma. For the cross-section, given a model, and provided the appropriate \texttt{UFO} and \texttt{param\_card} files are supplied, we calculate all the relevant tree-level amplitudes using the \texttt{standalone} subroutine within \texttt{MadGraph}~\cite{Alwall:2014hca}.

For definiteness, we use the scalar singlet DM model~\cite{Silveira:1985rk, McDonald:1993ex, Burgess:2000yq}. This expands the SM by including a real singlet scalar field, $s$, and a $\mathbb{Z}_2$ parity, where only $s$ is odd. The most general renormalizable scalar potential of the model is
\begin{equation}
    V = -\mu _H \, |H|^2 +  \lambda_H \, |H|^4 + \mu_s^2\,s^2 + \lambda_s\,s^4 + \lhs \,|H|^2\,s^2 ,
\end{equation}
where $H$ is the SM Higgs doublet. Due to the $\mathbb{Z}_2$ symmetry, $s$ does not acquire a vev and is stable. The mass of $s$, after electroweak symmetry breaking, is
\begin{equation}
    m^2_s = 2\,\mu_s^2+\frac{\lhs}{\lambda_H}\mu^2_H\,.
\end{equation}
The parameter $\lhs$ corresponds to the Higgs portal coupling, which is the only connection (besides gravity) between the dark and visible sectors. Finally, the DM quartic coupling $\lambda_s$ does not play a role in this analysis. 

The phenomenology of the singlet scalar model as a candidate for DM has been extensively explored in the literature, particularly when the Higgs portal is greater than $0.1$ ($\lhs \gtrsim 0.1$), allowing DM to have been thermally produced in the primordial Universe. In this scenario, the model can be tested in collider experiments~\cite{Barger:2007im, Kanemura:2011nm, Djouadi:2011aa, No:2013wsa, Craig:2014lda, Robens:2015gla, Han:2016gyy, Ruhdorfer:2019utl, Englert:2020gcp, Garcia-Abenza:2020xkk, Biekotter:2022ckj}, through direct and indirect DM searches~\cite{Yaguna:2008hd, Goudelis:2009zz, Profumo:2010kp, Urbano:2014hda, Duerr:2015mva, DiMauro:2023tho}, or by combining different experimental approaches~\cite{He:2009yd, Djouadi:2012zc, Damgaard:2013kva, Baek:2014jga, Curtin:2014jma, Feng:2014vea, Han:2015hda, Duerr:2015aka, Benito:2016kyp, GAMBIT:2018eea}. Specifically, combined analyses assuming standard cosmology~\cite{Cline:2013gha, Beniwal:2015sdl, GAMBIT:2017gge, Athron:2018ipf} suggest that the scalar $s$ could still serve as a viable thermal DM candidate, but only within a very narrow region of its parameter space. However, if the Higgs portal coupling is significantly smaller, around $\lhs \sim \mathcal{O}(10^{-11})$, the singlet scalar DM could have been produced through the nonthermal mechanism~\cite{Yaguna:2011qn, Yaguna:2011ei, Campbell:2015fra, Kang:2015aqa, Belanger:2018ccd, Bernal:2018kcw, Heeba:2018wtf, Huo:2019bjf, Lebedev:2019ton, Cosme:2023xpa}. Alternatively, the scalar singlet could have been produced thermally, if the DM quartic coupling is sizable $\lambda_s \sim \mathcal{O}(1)$, but from self-production and cannibalization reactions~\cite{Carlson:1992fn, Bernal:2015xba, Heikinheimo:2017ofk, Arcadi:2019oxh, Bernal:2020gzm}. We note that this model has also been studied in the framework of nonstandard cosmology~\cite{Yaguna:2011ei, Bernal:2018ins, Hardy:2018bph, Bernal:2018kcw, Allahverdi:2020bys, Silva-Malpartida:2023yks}, and in the context of Hawking evaporation of primordial black holes~\cite{Bernal:2020bjf}
 
Having specified the model, we have created the necessary \texttt{UFO} files using \texttt{SARAH}~\cite{Staub:2008uz, Staub:2012pb, Staub:2013tta}. Furthermore, the \texttt{param\_card} files have been generated through \texttt{SPheno}~\cite{Porod:2003um, Porod:2011nf, Staub:2011dp}. For the calculation of the cross section, since the squared matrix elements $\left| \mathcal{M} \right|^2$ depend only on the center-of-mass energy $\sqrt{s}$ and not on the solid angle, the integration of the phase space is straightforward, obtaining $\sigma(s)$ after combining all relevant processes.

\section{Impact of the EMD on DM abundance} \label{sec:DM_density}
The effect of having an EMD depends on whether the DM is generated via freeze-out (thermal) or freeze-in (non-thermal) mechanisms, and when exactly does the DM genesis happen. Let us first assume that DM production occurs at any point before Stage~4, that is, before reaching the temperature $\Tend$. For both freeze-in and freeze-out, posterior injection of the entropy from decay of $\phi$ suppresses the prediction of the present yield $Y_0$. Then, to reproduce the observed relic density, it is necessary to generate more DM than in the standard cosmological scenario. For freeze-in this means that one must increase the production cross section, which implies that the DM coupling to SM radiation $\lhs$ must be larger. For freeze-out, in contrast, one would need DM to decouple earlier, requiring a smaller coupling. 

\begin{figure}[t]
    \def\sepf{0.49}
    \centering
    \includegraphics[width=\sepf\columnwidth]{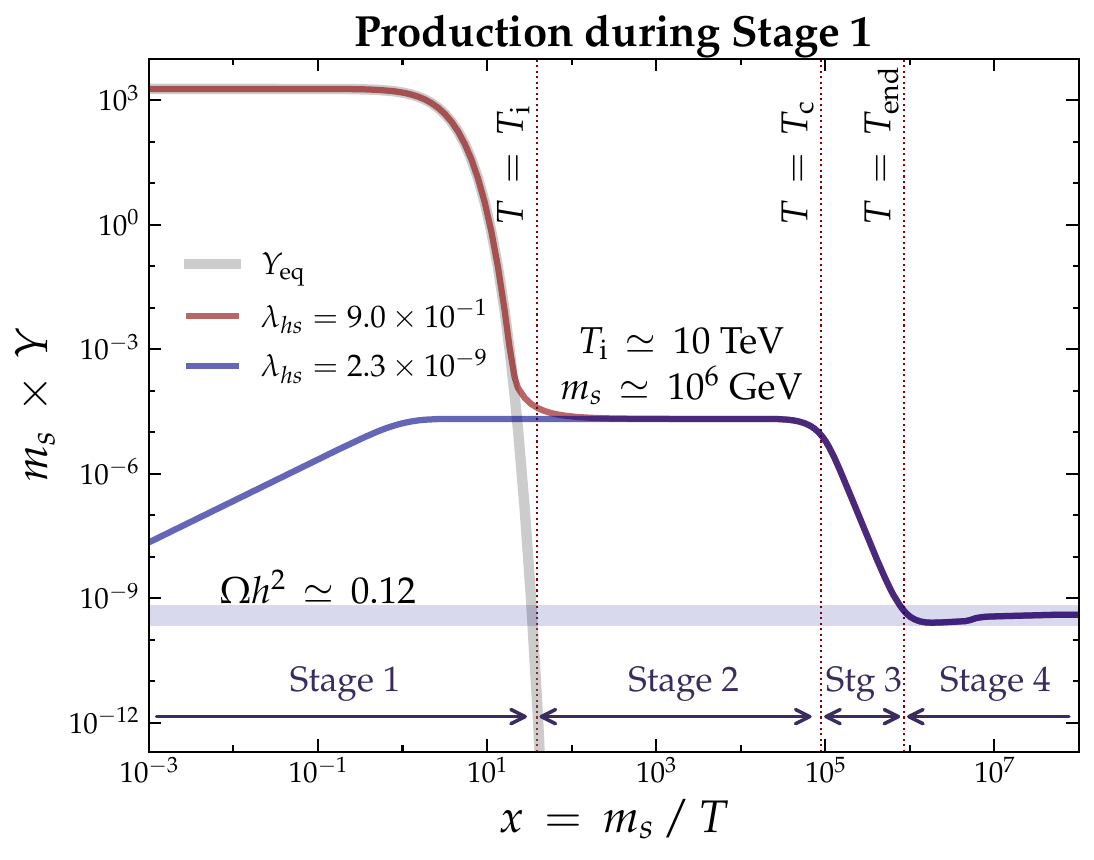}
    \includegraphics[width=\sepf\columnwidth]{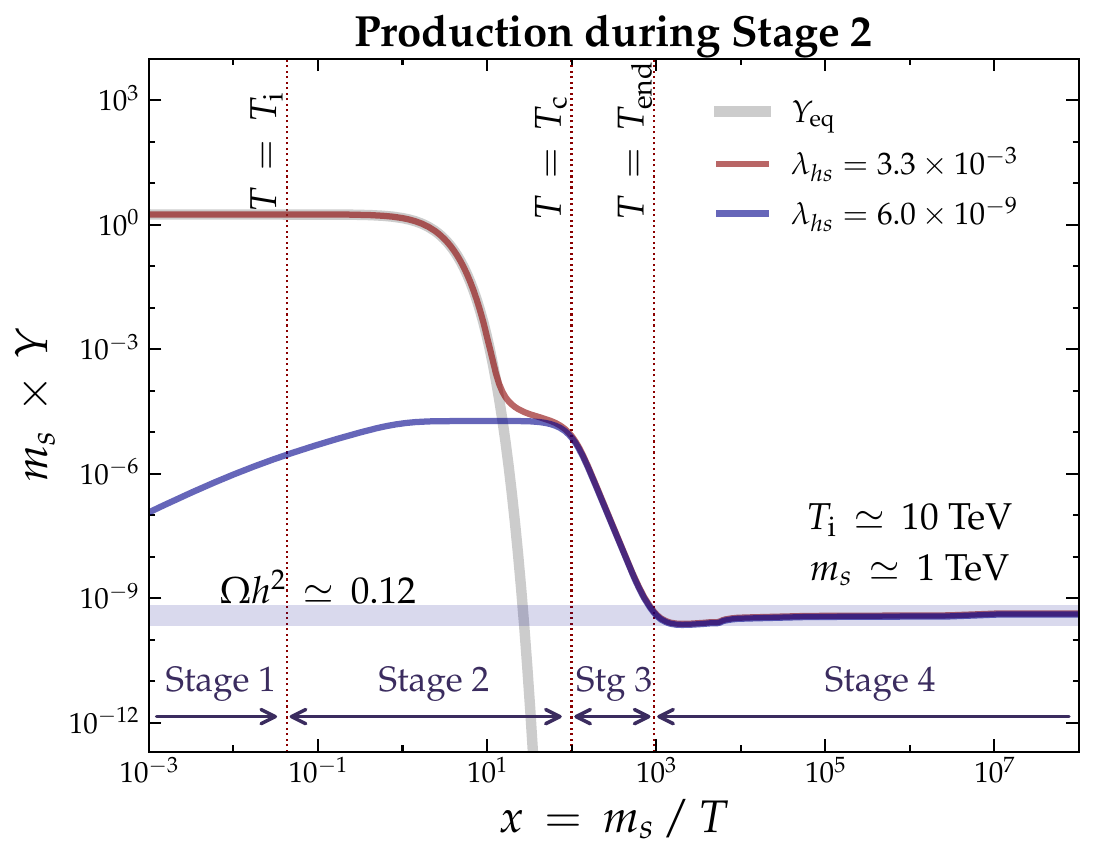}
    \includegraphics[width=\sepf\columnwidth]{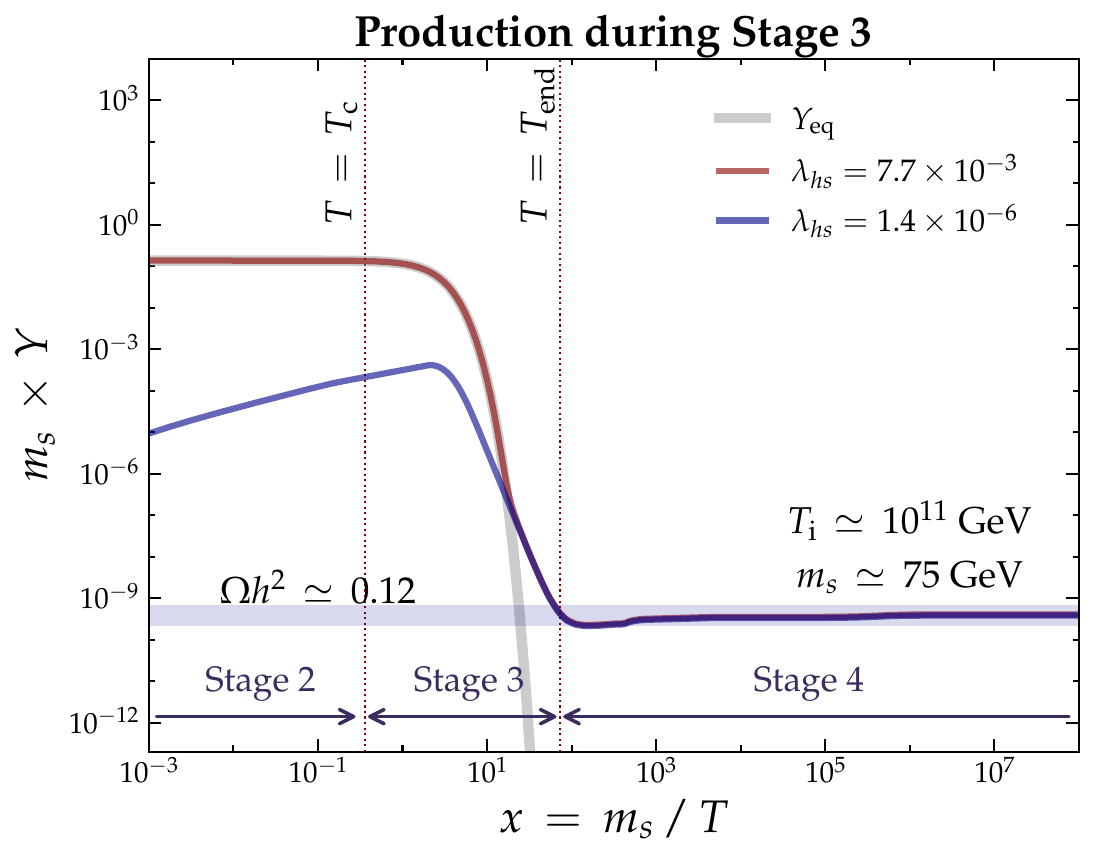}
    \includegraphics[width=\sepf\columnwidth]{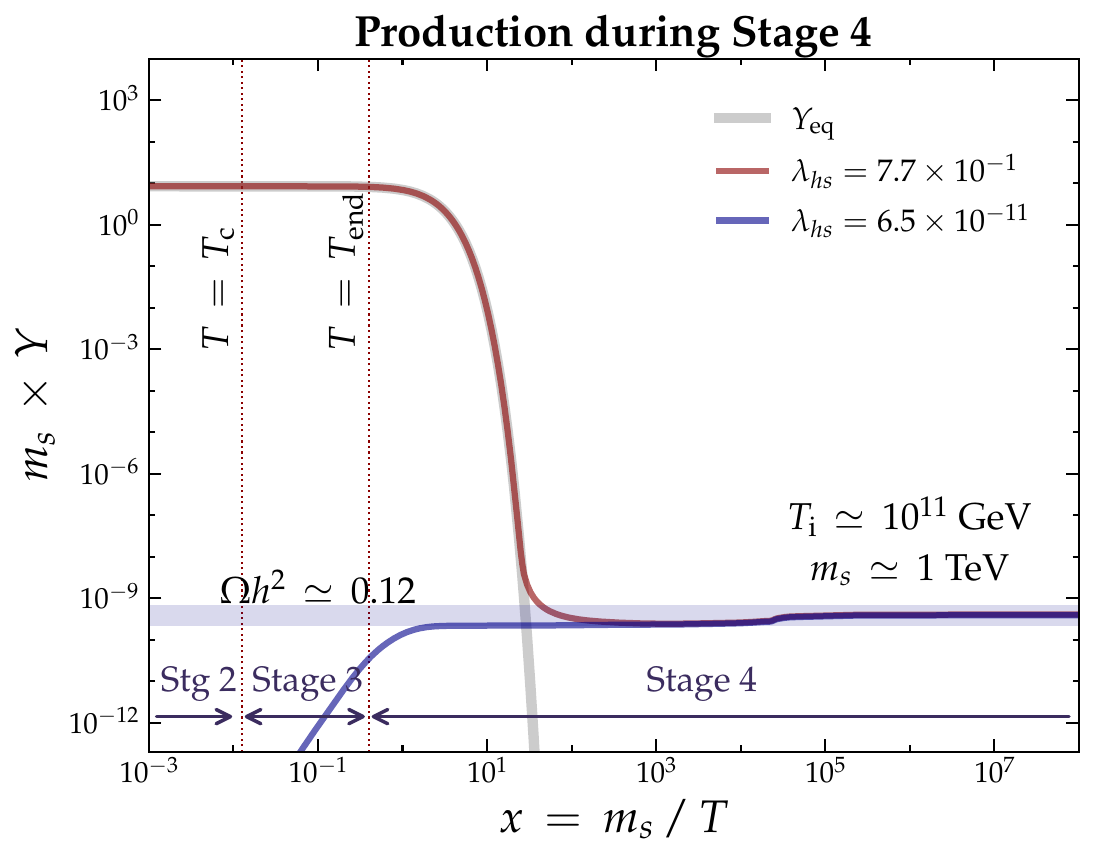}
    \caption{Evolution of the DM yield $Y$ as a function of the inverse of the SM temperature $T$ for $\Tend = 1$ GeV. The solid gray line represents the equilibrium yield $Y_{\rm eq}$, while the solid red and blue lines represent the DM yield for the thermal and nonthermal cases, respectively. The thick solid purple line shows the yield that reproduces the observed relic density.}
    \label{fig:yi}
\end{figure}

The picture described in the previous paragraph is sufficient to broadly understand our results. However, further intricacies depend on which stage of the evolution of the Universe DM was mainly produced. To this end, in Fig.~\ref{fig:yi} we show the evolution of the DM yield $Y$ relative to the inverse of temperature, for the freeze-in (blue) and freeze-out (red) scenarios that occur within each of the four stages described in Section~\ref{sec:EMD}. In all panels, the solid gray band shows the equilibrium yield, while the thick blue horizontal line represents the value of $\mdm\,Y$ corresponding to the observed relic density, $\Omega h^2 \simeq 0.12$. Furthermore, vertical dashed black lines denote temperatures $\Tini$, $\Tc$, and $\Tend$, associated with scale factors $\aini$, $\ac$, and $\aend$, respectively.

We start our description with the upper left panel, where DM either decouples from the SM radiation plasma or concludes its production within Stage~1, that is, before the $\phi$ field attains dominance ($T > \Tini$). Due to the very early DM genesis, the final value of the yield is not affected by the modified expansion rate of the Universe, being only altered by the decay of $\phi$ at Stage~3 ($\Tc > T > \Tend$). Thus, both thermal and non-thermal scenarios must obtain the same overabundance of DM at $T\sim \Tini$, which is then equally diluted by the maximal injection of entropy. It is interesting to note that the benchmark point used corresponds to a DM mass $\mdm \simeq 10^6$~GeV. In the standard cosmological scenario, masses larger than $\sim 130$~TeV are in tension with the unitarity bound~\cite{Griest:1989wd}, however, as can be seen here, larger masses (up to $\mdm \sim \mathcal{O}(10^{11})$~GeV) are allowed in non-standard cosmological scenarios~\cite{Berlin:2016vnh, Bhatia:2020itt, Bernal:2022wck, Bernal:2023ura, Coy:2024itg, Bernal:2024yhu}.

In the upper right panel, DM decouples or ends production during the Stage~2, at $\Tini > T > \Tc$. Here, the main difference is that the Hubble parameter is modified by EMD, as shown in Eq.~\eqref{eq:hub} and illustrated on the right panel of Fig.~\ref{fig:den1}. Given a fixed temperature, the Hubble parameter scaling as $a^{-3/2}$ instead of $a^{-2}$ would be larger than that in radiation dominance. This has consequences for both freeze-out and freeze-in scenarios.

For freeze-out, a larger Hubble rate means that the $\sv\, n_{\rm eq}/H$ ratio would be smaller. Then, Eq.~\eqref{eq:fo_cond} would be satisfied at larger temperatures, leading, in turn, to an earlier freeze-out and thus to a larger DM yield. This means that the necessary suppression of the value of $\lhs$ would not have to be as strong as expected from the injection of entropy alone.

To understand the effect of the modified Hubble parameter in freeze-in, it is useful to rewrite Eq.~\eqref{eq:boltzdm} in terms of the scale factor $a$ and $N \equiv n_s \times a^3$:
\begin{equation} \label{eq:approx_fimp}
    \frac{dN}{da} \simeq \frac{\sv}{a^4\, H}\, N_\text{eq}^2\,,
\end{equation}
with $N_\text{eq} \equiv n_\text{eq} \times a^3$. As can be seen, a different scaling of $H$ implies that $a^4 H \gg \sv N_{\rm eq}^2$ would be reached earlier (higher temperatures). Consequently, there is a lower yield at the peak of DM production. To reproduce the observed relic density, it is then necessary to increase the coupling beyond what is required by entropy injection.

In the lower left panel, DM ceases production within the temperature range $\Tc > T > \Tend$. This corresponds to Stage~3, where $\phi$ decays and injects radiation entropy, with $T$ now scaling as $a^{-3/8}$ instead of $a^{-1}$. It is important to note that even though both freeze-in and freeze-out depend on the radiation temperature, their predictions for the present yield do not depend on how $T$ scales with $a$. Thus, in this stage, the main modifications to DM production still come from entropy injection and the different scaling of $H$. However, the dilution of the yield is now applied after the end of the DM production, so it is not as strong as expected if $T \gtrsim \Tc$. This implies that the overabundance does not have to be as large as in the previous stages. For freeze-out (freeze-in) this means that the coupling must be larger (smaller) than in the previous stage.

Finally, the last scenario is shown in the lower right panel. In this case, DM ends production at Stage~4 ($T < \Tend$), marking a return to the standard case, with the present yield unaffected by the EMD era.

\begin{figure}[t]
    \centering
    \includegraphics[width=0.9\columnwidth]{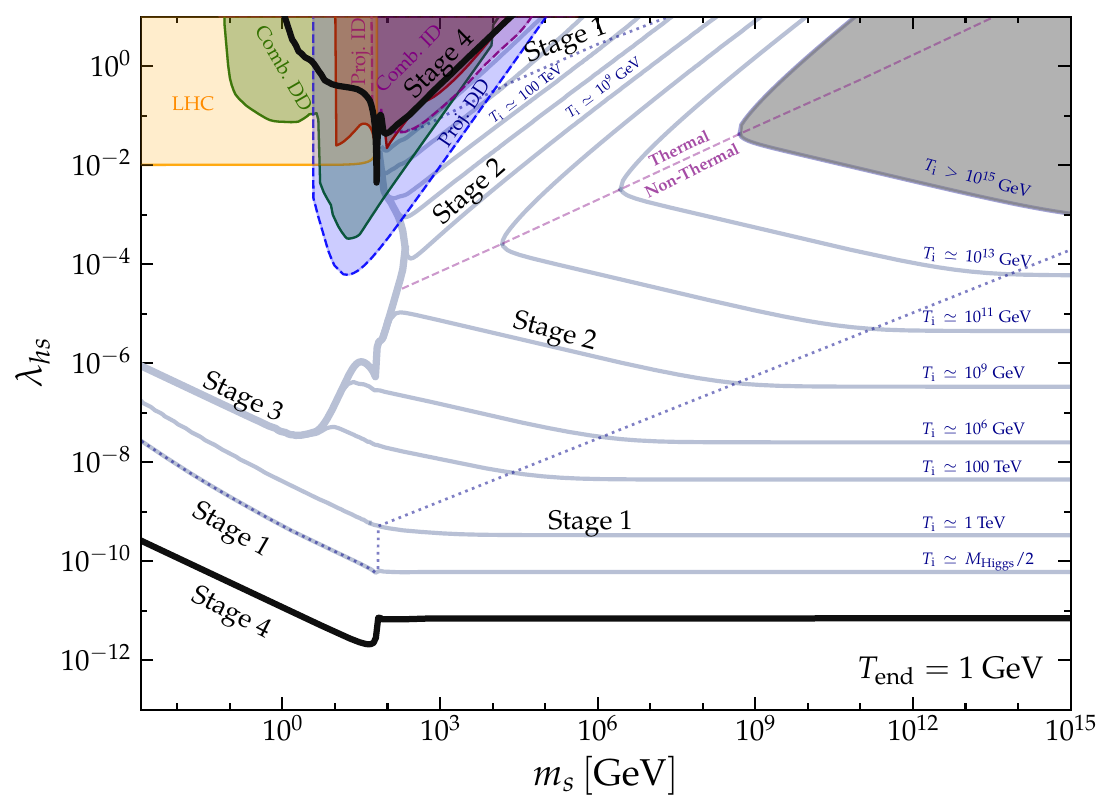}
    \caption{Parameter space of the model, for $\Tend=1$~GeV. The dashed blue lines reproduce the observed abundance of DM for DM production before $\Tc$, for different possible values of $\Tini$. The solid blue line corresponds to production after $\Tc$. The thick black lines correspond to standard cosmology. Constraints from Higgs invisible decays, direct and indirect detection, as well as future projections, are shown shaded on various colors.}
    \label{fig:1gev}
\end{figure} 
Let us now explore how these different behaviors are reflected in the parameter space of the model. Our results for $\Tend=1$~GeV are shown in Fig.~\ref{fig:1gev}, which corresponds to $\Gp \simeq 2\times10^{-17}$~GeV, and different values for $\Tini$. Here, we scan the DM mass $\mdm$ and the Higgs portal coupling $\lhs$, showing only perturbative couplings within a mass range of $1~{\rm MeV} < \mdm < 10^{15}$~GeV (for WIMPs, smaller masses pose challenges with BBN~\cite{Sabti:2019mhn, Sabti:2021reh}). The figure shows curves reproducing the present relic density, Eq.~\eqref{eq:presentY0}, which will be described in detail below. In addition, we show the constraint from the invisible Higgs decay~\cite{ATLAS:2022yvh, ATLAS:2023tkt, CMS:2023sdw}, in the yellow region, labeled ``LHC''. The green region represents a straightforward combination of these experiments, labeled ``Comb. DD''. This includes data from XENON1T~\cite{XENON:2018voc}, LUX-ZEPLIN (LZ)~\cite{LZ:2022lsv}, CDMSlite\cite{SuperCDMS:2015eex}, EDELWEISS~\cite{Lattaud:2022jnq}, and XENONnT~\cite{XENON:2023cxc}. Additionally, the blue regions, marked with the label ``Proj. DD'', display the projections of the XLZD consortium~\cite{Aalbers:2022dzr}. Finally, the red region, labeled ``Comb. ID'', combines various limits on the indirect detection of DM, including observations from MAGIC and Fermi-LAT~\cite{MAGIC:2016xys} as well as H.E.S.S.~\cite{HESS:2022ygk}. We also incorporate constraints from analyses based on CMB observations~\cite{Kawasaki:2021etm} and antiproton data from AMS-02~\cite{Cui:2018klo}. In the purple region, marked as ``Proj. ID'', we emphasize the projected limits for CTA~\cite{CTAConsortium:2017dvg} and SWGO~\cite{Viana:2021smp}.\footnote{The constraints of direct and indirect detection experiments on our parameter space is evaluated using \texttt{micrOMEGAs}~\cite{Belanger:2020gnr}.}

We first identify the curves leading to the correct $\Omega h^2$ in the freeze-out and freeze-in scenarios of standard cosmology, which appear as solid black lines at the top and bottom of the figure. These curves can also be associated with DM production occurring during Stage~4, that is, well after the EMD era. As is well known, the freeze-out mechanism for this model is essentially ruled out, while the freeze-in scenario presents major challenges, even for next-generation experiments. Furthermore, the maximal WIMP mass is $\mdm \sim 130$~TeV due to the unitarity bound~\cite{Griest:1989wd}.

In the figure, the present relic density is obtained on the gray lines, which are our main result. For different values of $\Tini$, the thin gray lines correspond to a DM production that occurs before $\Tc$, that is, during either Stage~1 or Stage~2, while the thick gray line corresponds to a production during Stage~3 ($\Tend<T<\Tc$). The figure also includes a dashed purple line, under which we have nonthermal production (freeze-in), that is, the $\sv\, n_{\rm eq}/H$ ratio is never greater than unity. Above this dashed line, the DM production is thermal (freeze-out), and the ratio is larger than unity at some point in the history of the early Universe.

We start our description under the dashed purple line, specifically, at very large masses and small couplings, corresponding to the freeze-in paradigm. In the lower right corner of the figure, we find that for the smallest evaluated $\lhs$ the present relic density can be obtained only if production happens during Stage~1. The reason for this is that we have a very large $T_{\rm fi} \sim \mdm$. If we were to have production in Stage~2, then $\Tini$ would be larger than $T_{\rm fi}$ (smaller $\aini$), which would imply a significant injection of entropy, as commented when discussing the left panel of Fig.~\ref{fig:den1}. This means that we would need to generate a considerable overabundance, which in turn can only be achieved by $\lhs$ much larger than those in this corner of the figure. In fact, for much larger couplings, one effectively finds that obtaining the observed $\Omega h^2$ in Stage~2 becomes possible. Similarly, one can argue that for a fixed $\lhs$, a smaller $\mdm$ implies a smaller $T_{\rm fi}$, such that at some point $T_{\rm fi}<\Tini$ without having a too large entropy injection, allowing DM production in Stage~2. 

The above arguments explain the blue dotted line that separates both stages, which has a constant slope as long as the DM is heavier than the Higgs ($\mdm\gtrsim M_{\rm Higgs}$). If $\mdm$ is smaller, then $T_{\rm fi}$ becomes independent of the mass of DM and is fixed around $M_{\rm Higgs}/2$. Thus, in the small-mass region of parameter space we find that we will have DM production in Stage~2 only if $\Tini>M_{\rm Higgs}/2$, leading to the line separating Stages~1 and~2 becoming vertical at some point, and then following the $\Tini \simeq M_{\rm Higgs}/2$ line.

Having understood how the Stages~1 and~2 production regions are defined, let us analyze the solid curves that lead to the correct relic density. For DM production during Stage~1, we find several values of $\Tini$ leading to the correct $\Omega h^2$, where for a larger $\Tini$ (larger injection of entropy) we need larger couplings. These are shown until a maximum value of $\Tini=10^{15}$~GeV, above which eventually $H \gtrsim H_I^{\rm CMB}$. Moreover, as in standard cosmology, once $\Tini$ is fixed, the required coupling is independent of $\mdm$. Thus, the lines for Stage~1 are parallel to those for standard cosmology. In contrast, for DM production during Stage~2, the $\phi$ field is no longer subdominant and has a significant effect on the evolution of the Hubble parameter. As explained in the discussion around Eq.~\eqref{eq:approx_fimp}, this requires an increase in $\lhs$ with respect to the one in Stage~1. Thus, in the region for Stage~2 we observe a change in the slope of the curves, which requires greater couplings as $\mdm$ and $T_{\rm fi}$ decrease.

On Stage~3, DM ceases its production within a temperature range $\Tc > T > \Tend$. The parameter space leading to the present relic density for this stage collapses to the thick gray line. As explained in Section~\ref{sec:DM_density}, two main effects are relevant: first, the change in the evolution of the Hubble parameter, and second, the smaller impact of the injection of entropy due to the decay of the field $\phi$. The net effect is that $\lhs$ does not have to be as large as in Stage~2. 

Interestingly, all the curves in Stage~2 merge with the single Stage~3 curve. The exact point where a Stage~2 curve merges occurs for $T_{\rm fi} \simeq \Tc$, with a specific $\mdm$ and $\lhs$. For the same $\Tini$ and smaller masses, we have $T_{\rm fi}<\Tc$, giving the lesser impact as aforementioned from entropy injection and allowing for a smaller coupling with respect to Stage~2. This means that a single point in Stage~3 can correspond to a specific $\Tini$ leading to $T_{\rm fi} \simeq \Tc$, or to any larger $\Tini$ corresponding to $T_{\rm fi}<\Tc$. Thus, unlike Stages~1 and~2, any point in Stage~3 becomes infinitely degenerate with respect to the possible values of $\Tini$ compatible with it, meaning that if $\mdm$ and $\lhs$ lie within this curve, it is not possible to derive any information regarding $\Tini$ and $\Tc$. Moreover, it is important to note that these effects in Stage~3 are consistent with the results presented in Ref.~\cite{Silva-Malpartida:2023yks}, which correspond to $\Tc \to \infty$. 

Regarding the region enclosed by the Stage~3 curve, we find no scenario capable of reproducing the observed relic density. The reason is straightforward: since the Stage~3 curve determines the coupling required to get the correct $\Omega h^2$, any coupling above it will generate an overabundance beyond the injection of entropy.

Finally, let us comment that in the figure the standard cosmology curve corresponds to a Stage~1 boundary. In particular, the Stage~1 curves are drawn towards the standard cosmology curve as $\Tini$ decreases. At the boundary, we actually have $\Tini\to \Tend$, meaning that the DM dilution becomes negligible and thus approaches a standard cosmology scenario. Notice that, strictly speaking, this does not correspond to the Stage~4 situation portrayed on the lower right panel of Fig.~\ref{fig:yi}, but is equivalent to it.

Similar arguments hold for the freeze-out scenario, which occurs above the dashed purple line. In this case, Stages~1, 2 and~3 always require suppressed couplings with respect to standard cosmology, with Stage~1 (Stage~3) requiring the largest (smallest) suppression. This drives similar changes in slopes as in the freeze-in scenario, albeit weaker. Interestingly, we emphasize that larger WIMP masses up to $\mdm \sim \mathcal{O}(10^{11})$~GeV are allowed in this case with an EMD era.

An interesting observation is that for very large $\Tini$ ($\Tini \gtrsim 10^{10}$~GeV) there exist non-thermal Stage~2 curves that do not merge into the corresponding Stage~3 curve, but rather merge with thermal Stage~2 curves. This implies that for these extreme values for $\Tini$, large (small) couplings needed to have $T_{\rm fi\backslash fo}=\Tc$ lead the non-thermal curves into the thermal regime and vice versa. However, this does not mean that such values of $\Tini$ do not allow DM to be produced within Stage~3, as commented earlier, for masses and couplings within the Stage~3 curve it is possible to have $T_{\rm fi\backslash fo}<\Tc$, which allows arbitrarily large values of $\Tini$. In other words, for very large $\Tini$ there exist two {\it disconnected} solutions that provide the present relic density: the thermal and nonthermal merging scenarios, and the disconnected Stage~3 curve.

With regard to the observability of this kind of framework, it is important to note that the present experimental techniques are already testing parts of the favored parameter space, while future direct- and indirect-detection experiments will be able to probe even larger regions of the parameter space. For this case, with $\Tend = 1$~GeV, future experiments can test regions corresponding to thermal production during all stages.

\begin{figure}[p]
    \centering
    \includegraphics[width=0.8\columnwidth]{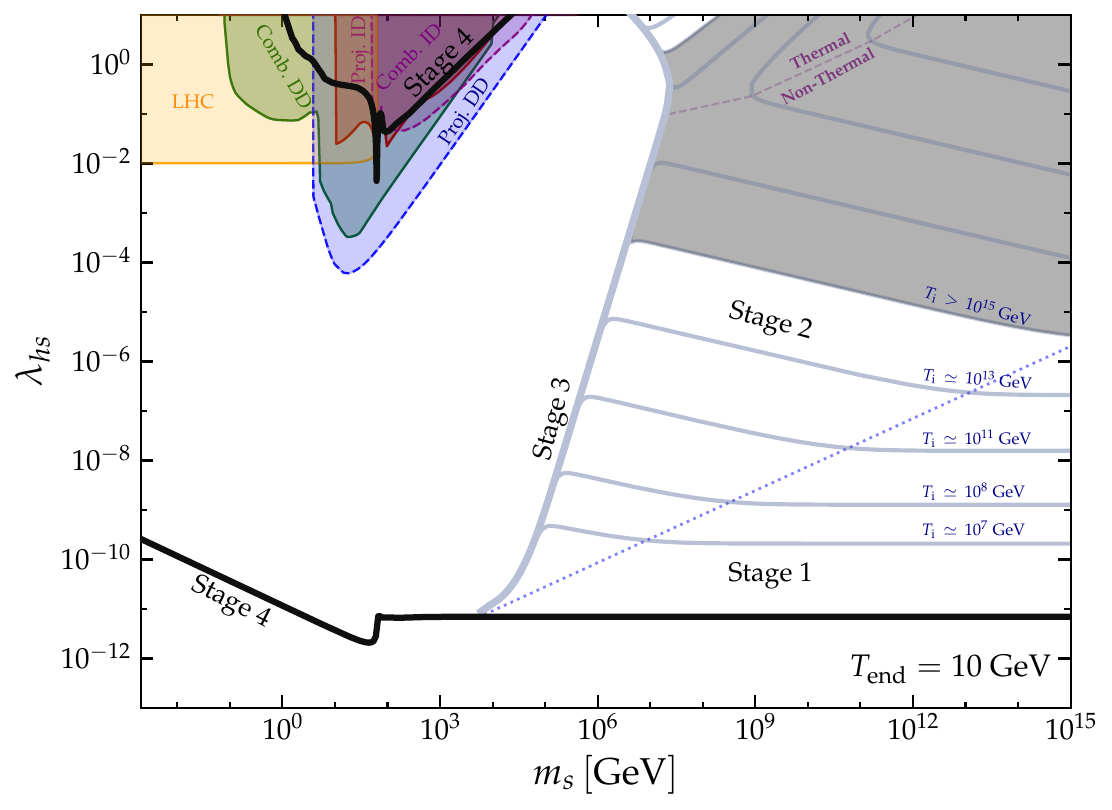}
    \caption{Same as Fig.~\ref{fig:1gev}, but for $\Tend=10\,$TeV.}
    \label{fig:10tev}
\end{figure}

\begin{figure}[p]
    \centering
    \includegraphics[width=0.8\columnwidth]{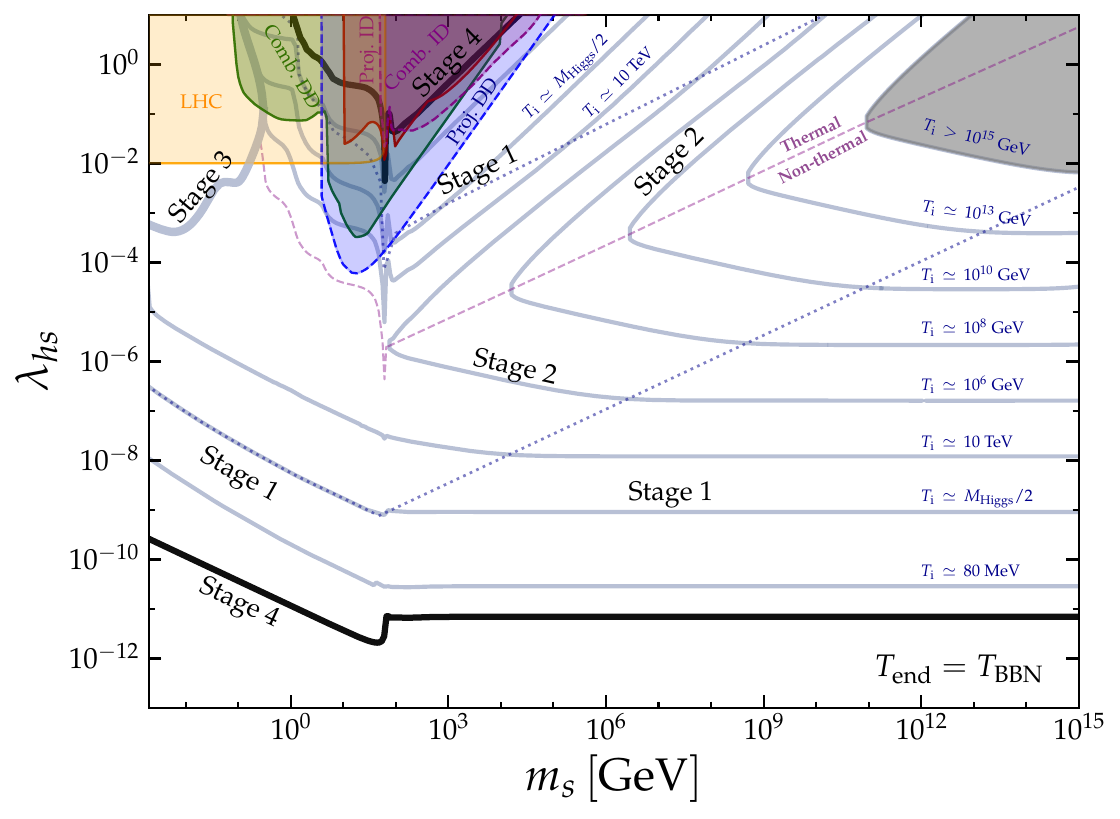}
    \caption{Same as Fig.~\ref{fig:1gev}, but for $\Tend = T_{\rm BBN}$.}
    \label{fig:4mev}
\end{figure}

Now that we understand the main consequences of the EMD scenario on the scalar singlet DM model, we study the consequences of varying $\Tend$. The results are shown in Figs.~\ref{fig:10tev} and~\ref{fig:4mev}, for $\Tend=10$~TeV and 4~MeV, respectively. As before, we will explain our results in terms of the freeze-in scenario since it occurs in a larger part of the parameter space.

If we increase $\Tend$ to 10~TeV (Fig.~\ref{fig:10tev}), we find that the Stage~1 and Stage~2 curves have a behavior similar to that of Fig.~\ref{fig:1gev}, but are now valid for higher values of $\Tini$. In fact, every point in these regions has an $\mdm$ and $\lhs$ combination implying a specific DM overproduction, which then has to be diluted by a specific amount of injection of entropy. Increasing $\Tend$ then requires an increment of $\Tini$, such that the difference in temperatures leads to a similar amount of entropy dilution. Furthermore, since we are now in a scenario where $\Tini$ is generally larger, it is possible to have DM production during Stage~2 ($T_{\rm fi}<\Tini$) for smaller values of $\lhs$, which lowers the blue dotted line separating Stages 1~and~2. Finally, having overall larger values of $\Tini$ implies that the shaded area ruled out by $H_I^{\rm CMB}$ will, of course, increase.

Similarly, increasing $\Tend$ leads to a shift to the right for the Stage~3 curve. This happens because no Stage~1-3 solutions are allowed for $T_{\rm fi}<\Tend$. Thus, no Stage~3 solution is allowed if $\mdm\sim T_{\rm fi}<\Tend$. For a smaller mass of DM, the only possibility is to have DM production during Stage~4, leading to a scenario similar to that shown on the lower right panel of Fig.~\ref{fig:yi}, that is, indistinguishable from standard cosmology.

In Fig.~\ref{fig:4mev} we decrease $\Tend$ to $T_{\rm BBN} = 4$~MeV, which is the lowest possible value of $\Tend$ consistent with BBN. Similarly to the previous case, with the decrease in $\Tend$ we also find an overall lowering of $\Tini$, shifting in turn the dotted line separating Stages~1 and 2 towards larger values of $\lhs$, and reducing the shaded region ruled out by $H_I^{\rm CMB}$. Since now we always have $\Tend<T_{\rm fi}$, the Stage~3 curve is shifted toward smaller values of the DM mass. It is interesting to note that, here again, regions of the favored by the EMD scenario are already in tension with direct and indirect detection data, and broader regions will be probed with next-generation experiments.

\section{Conclusions} \label{sec:concl}
Over the last few decades, significant experimental efforts have been made to understand DM, yet it remains one of the most challenging and fundamental issues in particle physics and cosmology. Concerning its origin in the early Universe, it is often assumed that DM is formed either thermally or non-thermally as a WIMP or a FIMP. From a cosmological standpoint, the common assumption is that the early Universe's energy density was dominated by radiation from the SM after inflationary reheating stopped, up until the era of BBN. Additionally, it is assumed that reheating concludes at a very high energy scale, leading to the production of DM when SM radiation dominates the Universe.

In this work, we have explored the impact of an early matter-dominated (EMD) era on DM genesis using the singlet-scalar DM (SSDM) model. By considering the evolution of a non-relativistic matter field $\phi$ alongside SM radiation, we identified four distinct stages in the Universe expansion history. In the initial stage ($T > \Tini$), SM radiation is the dominant component of the Universe. In the second stage ($\Tini > T > \Tc$), the field $\phi$ dominates, governing the behavior of the Hubble parameter. Throughout the third stage ($\Tc > T > \Tend$), $\phi$ still dominates but now efficiently decays, injecting entropy into the SM plasma. Finally, in the last stage ($T < \Tend$), SM radiation once again becomes the dominant factor in the evolution of the Universe. The effects of each stage on the DM yield can be compensated for by modifying the portal coupling, thereby expanding the parameter space that reproduces the observed relic density, as shown in Figs.~\ref{fig:1gev}, \ref{fig:10tev}, and~\ref{fig:4mev}. In particular, the WIMP paradigm can be realized with couplings much smaller than in the standard cosmological scenario, while much larger couplings are required in the FIMP case. This fact is particularly interesting, as it relaxes the experimental tension of WIMPs and increases the perspectives for a detection in the FIMP case. Furthermore, EMD eras help us to evade the unitarity limit of $\sim 130$~TeV for WIMP DM masses in the standard cosmological scenario, allowing masses up to $\sim 10^{10}$~GeV. Additionally, sizable regions of the favored by the EMD scenario are already in tension with direct and indirect detection data, and even broader regions will be probed with next-generation experiments.

For this paper, we used the SSDM model and explored the parameter space that reproduces the observed relic density and the smooth transition between the WIMP and FIMP paradigms. We used this model already implemented in \texttt{SARAH} and generated all allowed annihilations using \texttt{MadGraph}. Finally, we numerically solved the Boltzmann equation for DM. Although our results are derived from the SSDM model, the conclusions are expected to be applicable to a broader class of DM models.

\acknowledgments
JS and JJP acknowledge funding by the {\it Dirección de Gestión de la Investigación} at PUCP, through grant DGI-2021-C-0020. JS and RL acknowledge the support of the `Alianza del Pacífico' scholarship. NB received funding from the Spanish FEDER / MCIU-AEI under the grant FPA2017-84543-P. This project has received funding from the European Union’s Horizon Europe research and innovation programme under the Marie Skłodowska-Curie Staff Exchange grant agreement No 101086085 – ASYMMETRY.


\bibliographystyle{JHEP}
\bibliography{biblio}
\end{document}